\documentclass[12pt,usenatbib,iop]{emulateapj}
\newcommand       \micronm      {\,\mu{\rm m} }
\newcommand	      \nm           {\,{\rm nm}}

\newcommand       \sinvert		{\,{\rm s^{-1} }}
\newcommand       \flux             {\,$\rm erg\,s^{-1}\,cm^{-2}$ }

\newcommand       \surfbright              {\,$\rm erg\,s^{-1}\,cm^{-2}$\,arcsec$^{-2}$ }
\newcommand       \mh            {H$_2$}
\newcommand       \brg             {\,Br$\gamma$ } 

\newcommand       \HII          {\,H\,{\footnotesize II} }

\newcommand      \HeI       {He\,{\footnotesize I} }
\newcommand      \CO       {$^{12}$CO }

\begin{document}

\slugcomment{Accepted for publication in ApJ, April 18, 2015}

\title{Molecular and Ionized Hydrogen in 30 Doradus. I. Imaging Observations} 
\author{Sherry C. C. Yeh\altaffilmark{1,2}, Ernest R. Seaquist\altaffilmark{2}, Christopher D. Matzner\altaffilmark{2}, Eric W. Pellegrini\altaffilmark{3}}
\email{E-mail: yeh@naoj.org}

\altaffiltext{1}{Subaru Telescope, National Astronomical Observatory of Japan, 650 North A'ohoku Place, Hilo, HI 96720, USA}
\altaffiltext{2}{Department of Astronomy \& Astrophysics, University of Toronto, 50 St. George St., Toronto, ON M5S 3H4, Canada}
\altaffiltext{3}{Department of Physics and Astronomy, University of Toledo, 2801 West Bancroft Street, Toledo, OH 43606, USA}

\begin{abstract}
We present the first fully calibrated \mh\,\, 1--0 S(1) image of the entire 30 Doradus nebula. The observations were conducted using the NOAO Extremely Wide-Field Infrared Imager 
on the CTIO 4-meter Blanco Telescope.
Together with a NEWFIRM \brg image of 30 Doradus, our data reveal the morphologies
of the warm molecular gas and ionized gas in 30 Doradus. 
The brightest H$_2$-emitting area, which extends from the northeast to the southwest of R136, is a photodissociation 
region viewed face-on, while many clumps and pillar features located at the outer shells of 30 Doradus are photodissociation 
regions viewed edge-on. Based on the morphologies of \mh\,, \brg, \CO, and 8$\micronm$ emission,  the \mh\, to \brg line ratio and Cloudy models, 
  we find that the \mh\, emission is formed inside the photodissociation regions of 30 Doradus,  2 -- 3 pc to the 
 ionization front of the \HII region, in a relatively low-density environment $<$ 10$^4$ cm$^{-3}$. 
 Comparisons with \brg, 8 $\micronm$, and CO emission indicate that \mh\, emission is due to fluorescence, and provide no evidence for shock excited emission of this line.
\end{abstract}

\section{Introduction}
Starburst feedback is vital in galaxy evolution, as it is important in unbinding large molecular clouds \citep{krumholz06,fall10,murray10}, 
driving gravitational collapse inside molecular clouds and triggering sequential star formation \citep{elmegreen77,oey05,zavagno10}, 
driving turbulent motions within the clouds \citep{matzner02}, eroding molecular clouds by photo-evaporation \citep{whitworth79,williams97,matzner02}, 
and determining emission line spectra of the photoionized regions \citep{binette97,dopita05,dopita06,draine11,yeh12,yeh13,verdolini13}. 
While the effects of massive star feedback have been extensively 
discussed in the literature, 
a critical piece of information is missing: how does energy and momentum feedback from massive stars affect the molecular clouds' physical properties? 
We focus on the spatial distribution of molecular and ionized hydrogen emission, signals which highlight ionization fronts and working surfaces in this region of active stellar feedback.

A prime site for the origin of \mh\, emission is at the boundary of ionized gas, shocks, and
molecular clouds, which is the working surface for many forms of stellar
feedback into the dense gas. 
Because the excitation of \mh\, molecules is sensitive
to the density structure, radiation hardness, or mechanical energy input from shocks,
\mh\, ro-vibrational transitions are unique tracers to probe physical properties of the ISM under 
massive star feedback. 
The 30 Doradus nebula (30 Dor) in the LMC is a well-studied and one of the nearest (50 kpc) starburst regions, 
 at which distance a very high spatial linear resolution (1$\arcsec$ = 0.2 pc) can be achieved. The 30 Dor nebula is dominated by the very young star cluster R136, which produces 10$^{51.6}$ hydrogen-ionizing photons per second \citep{crowther98}, ionizing neutral material and driving it outward. 
Although the Dragonfish Nebula \citep{rahman11} and NGC 3603 \citep{conti04} are much closer ($<$ 10 kpc) and produce hydrogen-ionizng luminosities of 
 10$^{51.8}$ and 10$^{51.5} \sinvert$, respectively, 30 Dor's convenient location out of the galactic plane allows a straightforward comparison to 
 more distant starburst regions. 
Many observations of 30 Dor have been presented in the wavelengths from the X-ray to radio \citep{townsley06,chu94,poglitsch95,rubio98,indeb09,johansson98,indeb13},
 however, a fully calibrated \mh\, map of the entire 30 Dor has never been produced.  Wide field of view images of multiple ISM components are critical for followup studies on dynamics of the region, and for the selection of sites for targeted observations, such as high-resolution spectroscopy and interferometric studies, which necessarily involves smaller fields of view.

Observations of \mh\, in parts of 30 Dor have been done by \citet{poglitsch95} (P95) and \citet{rubio98} (R98).  P95 reported that the \mh\, 1--0 S(1) morphology 
appeared fragmented ($\sim$1$\arcsec$, 0.2 pc clumps), and suggested that the emission originated from dense molecular clumps. 
The P95 data is observed in a 3$\arcmin \times$3$\arcmin$ area in 30 Dor, which is a very small fraction of the region, while 
the R98 \mh\, data were collected in a small area ($<$ 2$\arcmin \times$2$\arcmin$) and not calibrated. 
 These observations were limited in sensitivity and field of view by the instruments available at the time, and 
the photodissociation region (PDR) physical conditions derived from P95 and R98 do not fully represent that in the entire 30 Dor.

In this paper, we present the first fully calibrated \mh\, 1--0 S(1) image as well as a \brg image of the full nebula, with a  1.0$\arcsec$ angular resolution.
We describe the observations and data reduction in \S~\ref{S:obs_reduction}. In \S~\ref{S:results}, we show the \mh\, and \brg morphologies and determine line ratios,
 identify areas of interest for further analysis, and 
investigate the spatial relations between \mh\, and \brg. In \S~\ref{S:cloudy_models}, we present photoionization models using Cloudy, in order to  constrain the
range of physical parameters inside the 30 Dor PDRs by comparing the modeled and observed \mh\, to \brg line ratio, and to explore the issue of bright line contamination we discovered during data reduction in the \mh\, image. 
The origin of the \mh\, 1--0 S(1) emission is  discussed in \S~\ref{S:discussion}.
Finally, we summarize the paper in \S~\ref{S:summary}.
In the paper, all \mh\, emission refers to the \mh\, 1--0 S(1) transition, unless indicated otherwise. \\

\section{Observations and Data Reduction}\label{S:obs_reduction}
We observed 30 Dor using the NOAO Extremely Wide-Field Infrared Imager (NEWFIRM; \cite{probst08}) on the CTIO 4-meter Blanco  Telescope, over three half-nights on November 10, 11, and 12, 2010. NEWFIRM has a field of view of 28$\arcmin \times$28$\arcmin$, and its pixel scale is 0.4$\arcsec$ per pixel. 
The \mh\, 1--0 S(1) (2.12$\micronm$) and \brg (2.17$\micronm$) emission line data were taken using the 2124 nm \mh\, and 2168 nm \brg narrow band filters, respectively.  See Table~\ref{tab:obs} for filter parameters. The broad-band continuum data were collected using the Ks filter. The total exposure time of the \mh\, image was 210 minutes, and 14 minutes for the \brg image. The observations were dithered in a random pattern in a 30$\arcsec$ box to fill the gaps between detector arrays.
The photometric standard star S121-E was observed in both 2124 nm and 2168 nm filters to serve as the flux calibrator. Table~\ref{tab:obs} summarizes details of the observations. 
Because the angular size of 30 Dor in the sky is about the same as the size of NEWFIRM field of view, in order to obtain sky images free of nebular emissions, we nodded the telescope on and off\footnote{The telescope was nodded off-target 1 degree north of R136.} the target following the sky-target-target-sky sequence. 

Data reduction was carried out using the NEWFIRM pipeline V1.3 \citep{swaters09}. Dark subtraction was first applied to the data, followed by a linearity correction and flat fielding. Sky background was determined by taking the median of four preceding and four subsequent off-source sky exposures which are free of extended emission, and the background level was scaled to match that in target images and then subtracted. An astrometric solution was obtained using the 2MASS catalogue, and all images were reprojected and stacked. The sky background was then redetermined and subtracted by masking objects 
(including stars and extended nebular emission) detected in the first pass, and new stacks of images were produced. \\

\begin{deluxetable*}{lccccc}
\tabletypesize{\scriptsize}
\tablewidth{0pt}
\tablecaption{30 Doradus Observation Summary}
\tablehead{
\colhead{Line} & 
\colhead{Vacuum Wavelength} &
\colhead{Filter FWHM ($\nm$)} & 
\colhead{Filter name} &
\colhead{Observed Dates\tablenotemark{a}} &
\colhead{Total Integration Time (minutes)}}
\startdata
Ks continuum & \nodata & 320.0 & Ks & Nov 10, 12 & 22 \\
\mh\, 1--0 S(1) & 2.121 $\micronm$ & 24.0 & 2124 $\nm$ &  Nov 10, 11, 12 & 210  \\
\brg & 2.166 $\micronm$ & 24.4 & 2168 $\nm$ &  Nov 11 & 14
\enddata
\tablenotetext{a}{Data were taken in year 2010.}
\label{tab:obs}
\end{deluxetable*}

\subsection{PSF Matching}
We carried out a photometric analysis of field stars using the software package SExtractor \citep{bertin96}. Stars with detection higher than 5$\sigma$  
in areas free of extended nebular emission are identified by SExtractor, and their photometric parameters, such as flux and FWHM, are recorded
in catalogs.
 We first extracted the mode of the distribution of FWHM in each stacked image, namely \brg, \mh\,, and Ks images, as the representative 
seeing in each image. It showed that \brg and Ks images have better seeing ($< 1.0\arcsec$) than the \mh\, image. 
We then convolved the \brg and Ks images with gaussian kernels until the convolved FWHM matches that of the \mh\, image. 
After the gaussian convolution, the seeing in \mh\,, \brg, and Ks images is 1.0$\arcsec$, and this is the resolution in every image presented in the paper 
unless indicated otherwise.\\

\subsection{Flux Calibration}
The standard star S121-E \citep{persson98} was observed as a flux standard.
 The \brg and \mh\, filters are centered closely at the Ks filter central wavelength, and we apply the 
magnitude-to-flux density conversion factors derived from the S121-E data, to flux calibrate the 30 Dor data. Data were taken under stable weather and nearly constant airmass, therefore the major uncertainties of measured fluxes come from bright emission line contamination in the Ks filter, which are discussed in \S~\ref{SSS:brightline}.

\subsection{Continuum Subtraction}\label{ss:contsub}
To produce emission line images, one must subtract continuum emission in the narrowband (NB) data.
The ideal way to subtract continuum in a NB image is to use an off-line center NB filter with identical FWHM, and 
such NB image shows continuum without contamination of any emission lines. 
However, off-line NB filters were not available when the observations were carried out, we therefore carried out continuum subtraction using the 
 broadband (BB) Ks data.
 
Mathematically, the observed fluxes in each filter (2.12$\micronm$, 2.16$\micronm$, and Ks) can be expressed as
\begin{eqnarray}
F_{2124} &=& F_{H_2} + F_{ct_{2124}}~~,  \nonumber \\
F_{2168} &=& F_{Br\gamma} + F_{ct_{2168}}~~,  \nonumber \\
F_{Ks} &=& F_{H_2} + F_{Br\gamma} + F_{uk} + F_{ct_{Ks}}~~, \nonumber
\end{eqnarray}

where F$_{2124}$, F$_{2168}$, and F$_{Ks}$ are fluxes measured in the 2.12$\micronm$, 2.16$\micronm$, and Ks filters, respectively. 
 The convention `ct' labels the continuum emission fluxes measured in a filter, and `uk' marks unknown emission line fluxes contained within the Ks filter. 

Continuum subtraction in \brg and \mh\, images then will produce:
\begin{eqnarray}
F_{2168} - \alpha F_{Ks}  
&=& F_{Br\gamma} + F_{ct_{2168}} \nonumber \\
&-& \alpha(F_{H_2} + F_{Br\gamma} + F_{uk} + F_{ct_{Ks}})~~,  \nonumber  \\
{\rm yielding} \nonumber \\
F_{Br\gamma} &=& \frac{F_{2168} - \alpha(F_{Ks} - F_{H2} - F_{uk})}{1-\alpha}~~, \label{eq:brg_contsub}		 
\end{eqnarray}

and 

\begin{eqnarray}
F_{2124} - \beta F_{Ks} 
&=& F_{H_2} + F_{ct_{2124}} \nonumber \\
&-& \beta(F_{H_2} + F_{Br\gamma} + F_{uk} + F_{ct_{Ks}})~~, \nonumber  \\
{\rm yielding} \nonumber \\
F_{H_2} &=& \left[\frac{F_{2124} - \beta (F_{Ks} - F_{Br\gamma})}{1-\beta}\right] + \frac{\beta}{1-\beta} F_{uk}~~, \label{eq:h2_contsub}
\end{eqnarray}

where  $\alpha F_{Ks}$ = $F_{ct_{2168}}$ and 
$\beta F_{Ks}$ = $F_{ct_{2124}}$. The equations are applicable at every pixel in the image.
 The values of $\alpha$ and $\beta$ in principle should be close to
 the FWHM ratio of the \brg to Ks filter and \mh\, to Ks filter, respectively.

We determine $\alpha$ and $\beta$ empirically by evaluating the stellar flux ratio of 2.16$\micronm$ to Ks, and 2.12$\micronm$ to Ks. 	
We employed SExtractor to extract stellar fluxes 
in the 2.12$\micronm$, 2.16$\micronm$, and Ks images. 
The scaling factor $\alpha$ and $\beta$ are 0.07 and 0.08, respectively, which is in good agreement with the filter FWHM ratios. 
The above equations are correct when the CCD response is linear in all three filters.
However, stars with counts $>$ 10,000 Analog/Digital Units are saturated, i.e. the CCD response becomes nonlinear, and they cannot be completely subtracted. Therefore we exclude these stars in the analysis. 

We found that the BB Ks data contain \brg, \mh\,, and possibly other emission lines, as well as the continuum emission,
which introduce contamination.
We evaluate and discuss the bright line contamination issue in \S~\ref{SSS:brightline}.

\subsubsection{Bright Emission Line Contamination}\label{SSS:brightline}
No direct information is available to us on the possible 
emission lines that,  other than \brg and \mh\,, might contribute to the Ks filter emission. 
However, \HeI is a likely candidate since it is observed in other star forming regions. 
For example, helium emission lines are 
reported in M16 in the 2$\micronm$ regime \cite[][hereafter L00]{levenson00}, in addition to \brg and \mh\, lines. 
Amongst the detected emission lines, \brg is the brightest in M16, and a bright \HeI line at 2.06$\micronm$ is 70\% of the total flux of  \brg.  
If the \HeI-to-\brg ratio is the same in 30 Dor as in M16, the contamination from \HeI in the Ks filter
 will be noticeable.

\brg emission line in 30 Dor is likely the brightest amongst the emission lines in the Ks filter. 
If the \HeI line has the same relative strength in 30 Dor as in M16 (70\%), 
Equation~\ref{eq:brg_contsub} shows that it is equivalent to about 5\% of the continuum-subtracted \brg emission, assuming it is distributed in the same way.
Therefore continuum subtraction in the \brg image may not be severely affected by bright emission line contamination, and 
Equation~\ref{eq:brg_contsub} can be approximated as 
\begin{eqnarray}
F_{Br\gamma} &\simeq& \frac{F_{2168} - \alpha F_{Ks}}{1-\alpha}~~. \label{eq:brg_contsub_corr}		
\end{eqnarray}

In the case of \mh\, continuum subtraction, however, bright emission line 
contamination becomes significant.  Following Equation~\ref{eq:h2_contsub}, the first-pass continuum subtracted \mh\, shows  
that if no correction was made, then there would be a
noticeable over-subtracted area.
Indeed, without correction, we noticed a region of negative emission representing the \brg emission in the vicinity of the nebula.
With careful visual inspection, we found that
the negative components well resemble the morphology of the highest surface brightness \brg emission. 
We then inspected the first-pass continuum subtracted \mh\, and \brg images on the pixel-to-pixel basis, comparing the pixel values 
of the negative component in the \mh\, image to the \brg image. There is a tight correlation between the
negative \mh\, pixels and brightest \brg pixels, confirming that the majority of over-subtraction comes from line emission which correlates strongly with \brg emission.

Let
\begin{eqnarray} 
F_{H_2}^{\prime}&\equiv& \frac{F_{2124} - \beta (F_{Ks} - F_{Br\gamma})}{1-\beta} ~~ {\rm and} \nonumber \\ 
\xi F_{Br\gamma} &\equiv& F_{uk} \nonumber~~,
\end{eqnarray}
Equation~\ref{eq:h2_contsub} can be rearranged as 
\begin{eqnarray} 
F_{H_2}^{\prime} &=& F_{H_2} - \frac{\beta}{1-\beta} \xi F_{Br\gamma}~~. \label{eq:h2_brightline}
\end{eqnarray}

The slope $\frac{\beta}{1-\beta} \xi$ in Equation~\ref{eq:h2_brightline} is -0.10,  evaluated from fitting the residual data. Therefore we correct 
continuum subtraction of the \mh\, image by 
\begin{eqnarray}
F_{H_2}^{\prime} &=& F_{H_2} - (-0.10)F_{Br\gamma} \nonumber 
\end{eqnarray}

We inspected the pixel values of the corrected \mh\, image $F_{H_2}^{\prime}$, and the majority of the negative pixels after the correction have values around 0. 
In fact the factor 0.10 is higher than the expected value $\beta$/(1-$\beta$)$=$ 0.087, which indicates that we have not only corrected for the contamination introduced by
the \brg emission, but also the contamination from other unknown bright emission lines. The bright-line contamination is thus (0.10-0.087)/0.087 = 0.15, or 15\% of the \brg emission flux, which corresponds to 1.3\% of the continuum flux.

We found that, after correcting the \mh\, image for the contamination, some negative pixels still persist in the areas very close to R136. 
Those pixels can only be corrected by $\frac{\beta}{1-\beta} \xi$ = 0.14, instead of 0.10 in Equation~\ref{eq:h2_brightline}.
However this leads to over-correction in the image, i.e. bright \brg features become prominent in the \mh\, image, which
indicates that the contaminating emission is stronger relative to \brg near the cluster than further away.
The corresponding total brightness of this contamination is 61\% of the \brg flux, 
assuming that the emission line flux is distributed in the Ks band in the same way as \brg,
which leads to additional 4\% of the continuum flux, which is also the uncertainty in the contamination-corrected 
\mh\, image.
We suspect that the \HeI line at 2.06$\micronm$ is the major source of continuum contamination other than \brg, and its contamination becomes more 
significant in the central region of 30 Dor.
Several \HeI lines are detected in the M16 \HII region, in addition to \brg and \mh\, 1--0 S(1) (L00). 
As noted earlier, the brightest 
He emission line detected in M16 in the K band is \HeI at 2.06$\micronm$, and its flux of the \HeI line 
is 70\% of the total flux of \brg. The 61\% of the \brg flux contamination we have 
empirically estimated for the region near R136 is as significant as that in the M16 case.  
 Note that variations in the continuum slope, such as those induced by variations in extinction, will also not be consistent with constant values of our $\alpha$ and $\beta$ parameters.

We do not have any He line data in 30 Doradus to constrain the degree of contamination, therefore we turn to 
Cloudy simulations to explore this issue, which is discussed in \S~\ref{S:cloudy_models}. 
In this paper, the \mh\, image is empirically corrected for contamination following Equation~\ref{eq:h2_brightline}.

\section{Results}\label{S:results}

\subsection{\mh\, and \brg Morphologies}\label{SS:morphology}

\begin{figure}
\epsscale{1}
\plotone{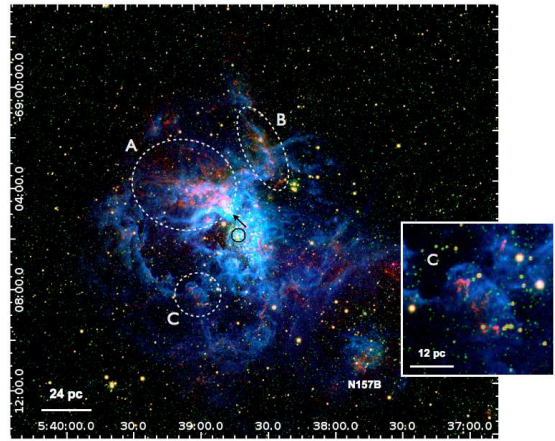}
\begin{center}
\caption{Three-color composite image of 30 Doradus. North is up and East is to the left. Red: \mh\,, displayed in linear scale; green:  Ks continuum in logarithm stretch; blue: \brg in logarithm stretch. R136 is marked by a black circle.
The seeing is 1.0$\arcsec$.
\mh\,-emitting areas of interest are marked by white dashed ellipses, including 
(A) the bright \mh\, band to the northeast of R136, (B) filaments pointing north-ward of R136, and (C) pillars to the southeast of R136. A zoomed-in figure of Area C is shown on the right.  A bright \mh\, finger in Area A is indicated by a black arrow.
}
\label{fig:3color-scale}
\end{center}
\end{figure}

Fully calibrated \mh\, and \brg images of 30 Dor are presented in a 3-color composite image in Figure~\ref{fig:3color-scale}. Red is \mh\,, blue is \brg, and green is the Ks band continuum.
Both \brg and Ks are stretched logarithmically to emphasize areas of highest surface brightness, while \mh\, is displayed in the linear scale because the line brightness dynamic range is much smaller. The star cluster R136 is marked by a black circle.  Note that the pulsar wind nebula N157B, which is located at the same distance  and 7$\arcmin$ to the southwest 
of 30 Dor, is also present in the image. In this paper we focus only on the analysis of 30 Dor and will ignore N157B. 

The \brg emission reveals the spatial distribution of ionized gas in 30 Dor. Areas with highest surface brightness appear to trace 
an arch structure extending from northeast to southwest of the R136 cluster, consistent with the ionized gas morphology reported in the literature  \citep{chu94,poglitsch95,pellegrini10}. 
To the north and west of R136, \brg appears quite filamentary and its surface brightness becomes lower. To the southeast of 30 Dor, the \brg morphology reveals
multiple shell structures, enveloping pillars and clumpy features.  
The total flux in an area 3.1$\arcmin \times$ 3.7$\arcmin$ is measured as 1.02$\times$10$^{-10}$ \flux. Although this is higher than 
 the total \brg flux reported in P95 (4$\times$10$^{-11}$\flux) measured in the same area, our \brg image detects fainter
 structure than that in the P95 image. 

The most prominent \mh\, emission is seen to the northeast of R136, in conjunction with the bright \brg arch and coincident 
with lower surface brightness \brg emission. This prominent \mh\, emitting area, which is also known as 30Dor-10 \cite[e.g.][]{johansson98,indeb09,indeb13},  also marked as Area A in Fig.~\ref{fig:3color-scale}, spans at least 3$\arcmin$ by 3$\arcmin$ in the plane of the sky, which corresponds to 
36$\times$36 pc at a distance of 50 kpc. 
The \mh\, emission in 30Dor-10 seems somewhat disordered and extended, with clumps close to R136 and filaments extending away from R136. 
The total \mh\, flux measured in the area is  5.16$\times$10$^{-12}$\flux after masking bright saturated stars.
The \mh\, emission to the west of R136 appears much clumpier and mixed with high surface brightness \brg emission.  
To the north and northeast of R136  (Area B), \mh\, appears in the form of filaments which seem to form a chimney pointing away from R136, and their morphology is 
poorly correlated with the \brg filaments in the same area. Prominent pillar features 
are seen in the southeast of R136, pointing towards the ionizing source R136 and are encompassed by the \brg emission,  such as that in Area C. 
These \brg envelopes have sharp outer boundaries, with radii of curvature significantly greater than those of the \mh\,  pillars they envelop.  This is suggestive of a photo-evaporative flow bounded by the pressure of hot gas.
No \mh\, emission was detected to the northwest of R136. 
 Overall, the \brg and \mh\, emission appear to form walls of the cavities or holes seen in 30 Dor. The observed \mh\, emission is located  well within the nebula (defined by optical broadband data). 
  We will discuss Area A, B, and C in detail in \S~\ref{SS:area_interest}.

\subsection{Areas of Interest}\label{SS:area_interest}
\begin{figure}
\begin{center}
\plotone{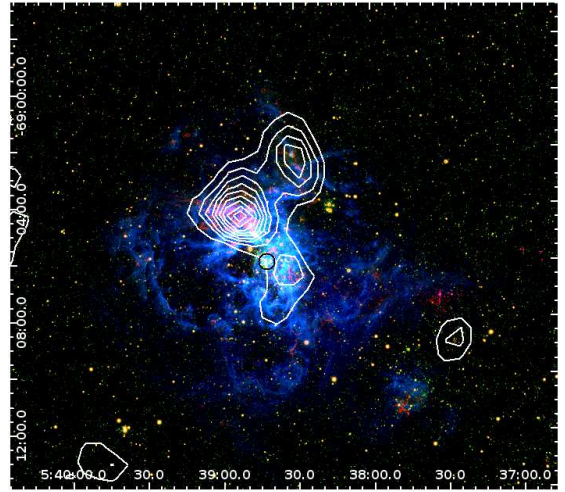}
\epsscale{1}
\caption{ CO 1--0 contours (white contours, \citealt{wong11}) superimposed on the 3-color composite image as shown in Figure 1.  
The CO contours
start at 10\% of the peak intensity (12.4 K km s$^{-1}$) and increase in 10\% steps.
The CO data angular resolution is 60$\arcsec$.
R136 is marked by a black circle.}
\label{fig:3color-co}
\end{center}
\end{figure}

We identified areas of interest for further analysis by comparing the \mh\,, \brg, and \CO morphologies as shown in Figure~\ref{fig:3color-co} and Figure~\ref{fig:3color-scale}.
 The \mh\, and \brg data were superimposed with the \CO 1--0 data of 30 Dor \citep{wong11}\footnote{We chose the \citet{wong11} CO data for comparison instead of the \citet{indeb13} ALMA data, because we are interested in the CO distribution in the entire 30 Dor. The ALMA image field of view is too small, and the interferometric data is not sensitive to extended structures.  Nevertheless, the ALMA data will be useful for detailed studies of smaller objects of interest in the region.}. The \brg image in Figure~\ref{fig:3color-co} is displayed in grayscale to guide the reader's eyes 
on the ionized gas morphology, while the \mh\, contours are convolved to a 4.5$\arcsec$ resolution. 
We found that bright H$_2$-emitting areas spatially correlate very well with the bulk of CO emission (Figure~\ref{fig:3color-co}). Based on the morphological 
correlation between \mh\, and CO emission, and \mh\, and \brg emission, three areas are identified as areas of interest:
(1) Area A: the northeastern band of \mh\, emission with highest \mh\, surface brightness, and it spatially coincides with the 
brightest CO emission in 30 Dor. This area is also known as 30Dor-10 as noted earlier;
(2) Area B:  the filament pointing north and away from R136,  which is relatively bright and spatially correlated well with the peak of CO emission in the 
same area; and 
(3) Area C: pillars to the southeast of R136 at the outer shell of 30 Dor, which have clearly defined
morphology in the \mh\, emission and are surrounded by Br$\gamma$-emitting envelopes; no CO emission is detected in this area  (by the ATNF Mopra Telescope, \citet{wong11}). A zoomed-in figure of Area C is
shown in Figure~\ref{fig:3color-scale}. 

 \mh\, was detected in an earlier observation (P95) in the western part of Area A, in a much smaller 2$\arcmin \times$ 2$\arcmin$ area. 
The \mh\, morphology observed by P95 was reported very clumpy, 
which is consistent with the \mh\, morphology shown here (Figure~\ref{fig:3color-scale}). 
The peak \mh\, surface brightness in the P95 result is 5.41$\times$10$^{-15}$\surfbright.
However, this peak intensity arises from a bright star in the field thus may not be reliable. 
In our wide-field observations, more \mh\, emission is seen towards the eastern part of Area A. 
After excluding bright saturated stars, the maximum \mh\, surface brightness is measured 2.15$\times$10$^{-15}$\surfbright.
 \mh\, in Area A also coincides with the bulk of CO emission in 30 Dor \citep{wong11} (Figure~\ref{fig:3color-co}). 
No \mh\, emission is seen completely uncorrelated with either the \brg or \CO 2--1 emission.
With only one ro-vibrational transition of \mh\, emission, and without making further 
assumptions, we estimated mass contained in the upper state of the 
observed \mh\, transition in Area A to be 0.01 M$_\sun$,  estimated with the Einstein A coefficient 
$A_{S(1)}$=2.09$\times$10$^{-7}$ s$^{-1}$ \citep{turner77,wol98}.
\footnote{Note that this mass estimate yields a lower limit of molecular mass, because 
it does not include the mass associated with all other levels.}
 This mass is orders of magnitude lower than the molecular mass estimated from CO emissions  of 8.5$\times$10$^4$ M$_\sun$ \citep{pineda12,johansson98}, 
 which implies that the observed \mh\, emission requires the existence of only a negligible fraction of the total molecular mass.

Area B also shows high \mh\, intensity, and the maximum \mh\, surface brightness in this area is 6.44$\times$10$^{-16}$\surfbright.
This region was outside the field of view of the P95 observations.
Area B lines up well with its CO counterpart (Figure~\ref{fig:3color-co}), similar to Area A.

Area C was outside the field of view of the P95 observations as well. The peak surface brightness is measured 1.76$\times$10$^{-15}$\surfbright, comparable 
to that measured in other areas. 
Although Area C does not appear to have any CO counterpart, its \mh\, emission indicates
the presence of molecular clouds in that area.
The morphology of the pillars is very similar to the ones observed in smaller \HII regions, such as M16 \citep[][L00]{hester96}. 
 We highlight few interesting morphological features in 
 Area C, which will lead to followup detailed studies of the region: (1) the \mh\, emission is enclosed within the \brg emission which extends toward R136, (2) 
 the \mh\, emission appears at the leading edge of regions where other molecular cloud tracers are present, such as the 8$\micronm$ emission, and (3) 
 the \brg emission is bounded by a sharp outer boundary whose radius of curvature is significantly larger than that of the \mh\,, which implies 
 possible pressure confinement by hot gas. 
A crude estimate of the projected distance between the \brg envelopes and the pillars is $\sim$ 3 pc. Area C's neighbor clumps  
also show similar separation between \brg and \mh\,. We will return to discuss this point in \S~\ref{SS:h2_density}.

\subsection{The Spatial Relationships Between Ionized and Molecular Gas in 30 Doradus}
The overall morphology of the observed \mh\, emission is well correlated with the \brg distribution within 30 Dor -- when \brg emission is present, 
\mh\, emission is either seen very nearby the \brg emission, such as the outer shells and pillar features, or appears to be a counterpart of the \brg emission,
such as Area A and the region just west of R136.

 Bright H$_2$-emission is seen to be spatially coincident with the \brg emission  in Area A (Figure~\ref{fig:3color-scale}), except for an excess of \brg emission around the limb of the ``finger" seen in \mh\,. 
This suggests that we are viewing the \HII region roughly face-on, except around the limb.    Furthermore, the CO-emitting gas in the core of the finger is likely to be marginally optically thick at 2.2\,$\mu$m, judging from the ALMA observations of $^{13}$CO presented by \citet{indeb13}.   They estimate $^{13}$CO column densities $>5\times10^{15}$\,cm$^{-2}$ along this finger (and clumps an order of magnitude higher column; their Fig.~10),  and adopt $N_{\rm H_2}=5\times 10^6 N_{\rm ^{13}CO}$; along with a dust opacity of roughly $1.3\times10^{-20}\,\rm cm^2$ per H atom at 2.17\,$\mu$m \citep{weingartner01}, we estimate the characteristic optical depth of the finger to be $\sim 0.65$.    It is therefore quite likely that most of the observed emission is emitted on the Earthward side of the molecular finger.  The absence of obvious shadows associated with high-column clumps of $^{13}$CO corroborates this interpretation.

On a similar basis, we infer that emission seen just west of R136 is likely to originate from behind the cluster (as seen from Earth), on the basis that the \citet{wong11} CO contours extend into this region. 
This geometry is further supported by recent optical emission line studies by \citet{pellegrini11}(P11). With optical emission line ratios together with
modeled ionization parameters, P11 suggests that Area A and the region just west of 30 Dor are located 60 pc behind R136. 

Area B may share the same geometry as the regions we discussed above -- this filament of molecular gas should be behind the \brg emission and R136. 
Although the \brg morphology is filamentary with rather poor correlation with the \mh\, emission, \brg is seen to partially coincide with the \mh\, and CO emission in the area, 
instead of being on the edge of molecular clouds. This suggests
that the \brg emission is not shielded by molecular clouds, and 
we are viewing this region face-on.  Lacking high-resolution CO data, however, we cannot draw this conclusion with certainty.

Area C is likely 
 oriented differently relative to 30 Dor than are Areas A and B. 
The bright \brg emission in this region shows relatively poor spatial correlation with the \mh\, emission. \brg is seen closer to R136 in the projected plane of sky, 
and it envelops the \mh\, emission,  which has a relatively small projected area. 
Although \brg extends outward from the \mh\, emission (and toward 30 Dor) to a similar degree here as in Areas A and B, the much smaller \mh-emitting regions give it more of the appearance of an evaporation flow. 

 In all three regions there are locations where \brg emission extends away from a boundary of the \mh\, emission, generally in the direction toward 30 Dor, before terminating at an enclosing boundary about two to three parsecs away.  To be specific, in Figure \ref{fig:3color-scale} we measure projected standoff distances of 2.5\,pc perpendicular to the finger in Area A (to the southeast); of 2.2\,pc (toward 30 Dor) from the small \mh\, blob at the southern end of Area A; of about 1.9\,pc (toward 30 Dor) from parts of the \mh\, ridge in Area B; and of 1.9\,pc (toward 30 Dor) and 2.2\,pc (perpendicularly to the northeast) from the \mh\, pillars in Area C.  All of these characteristics -- the tendency of \brg emission to enclose \mh\,, its tendency to extend a few parsecs toward 30 Dor, and its tendency to meet a sharp boundary, are expected to reflect the physical origin of these two types of emission, as well as the relative importance of effects such as photo-evaporation and confinement by stellar wind pressure.

\subsection{\mh\, to \brg Line Ratio}\label{SS:h2_to_brg_ratios}
The morphological relations between \mh\,, \brg, and CO emission are indicative that the observed \mh\, emission traces the PDRs in 30 Dor.
With data of only one \mh\, emission line, we do not have sufficient information to firmly constrain the physical quantities in the PDRs.
Nevertheless, the line ratio of \mh\, to \brg is a useful guide to delineate the spatial distribution and structure of molecular gas relative to ionized gas in the PDRs,
especially when a PDR is viewed face-on. Combining with numerical modeling efforts, the observed \mh\, to \brg line ratio offers a hint of the 
physical properties of molecular and ionized gas in 30 Dor.

\begin{figure}
\begin{center}
\plotone{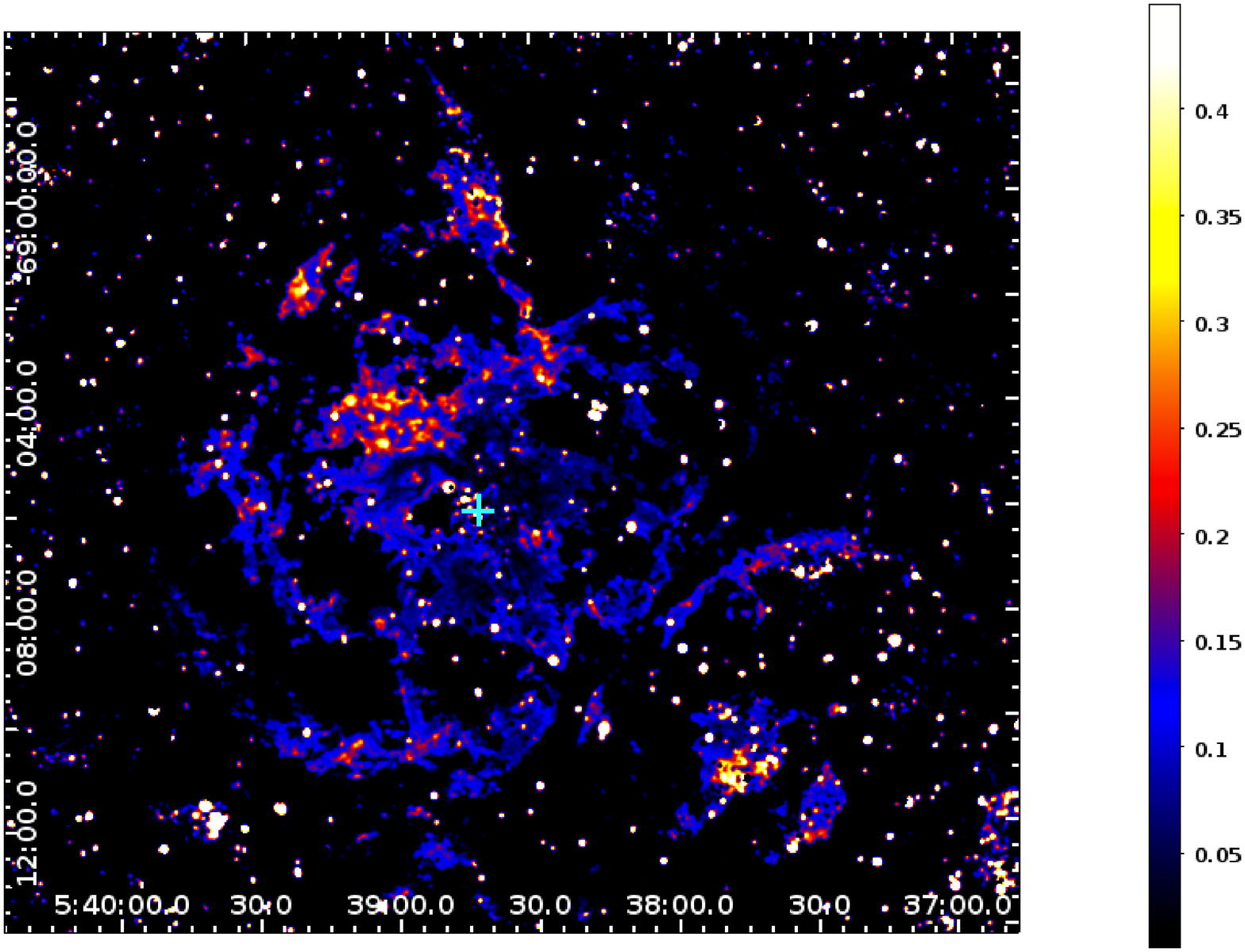}
\epsscale{1}
\caption[\mh\, to \brg line ratio map in 30 Dor]{ \mh\, to \brg line ratio map of 30 Dor, convolved by Gaussian kernels to a 4.5$\arcsec$ resolution. The colorbar marks the \mh\, to \brg ratio; R136 is marked by a cyan cross.}
\label{fig:h2_to_brg}
\end{center}
\end{figure}

A line ratio map \mh\, to \brg is shown in Figure~\ref{fig:h2_to_brg}. 
Pixels in the \brg image were clipped at a 3$\sigma$ level, and pixels with S/N higher than 50$\sigma$\footnote{An empirical value to mask saturated bright stars.} detection in the \mh\, image were masked 
in order to exclude bright saturated stars. A line ratio map then was convolved to a 4.5$\arcsec$ resolution with a gaussian kernel.

Overall, areas with higher \mh\,/~\brg
ratios in 30 Dor are clumpy with ratios of 0.2 to 0.5. Most areas show lower \mh\,/~\brg ratio $\sim$0.1, which agrees with the same line  
ratio observed in M16 (L00, derived from the total fluxes). 
Higher \mh\,/~\brg ratios are seen in Area A, Area B, filaments north of Area B, and some isolated pillar features (including Area C)
at the outer shells of 30 Dor. 

In Area A, the \mh\, to \brg ratio across the area appears clumpy, with some localized high line ratio areas and 
`voids'. The maximum ratio is 0.5, and the overall ratio is $>$ 0.2. 
The clumpy \mh\, distribution and \mh\, to \brg line ratios seen in Area A indicate that FUV radiation could penetrate deeper into the 
molecular clouds.
The maximum line ratio at Area B is 0.45, and the overall radio is $\sim$ 0.3. 
The line ratio morphology suggests that \brg across this area is fairly filamentary. 
Area C displayed a high \mh\, to \brg line ratio of 0.3, which coincides with the bright H$_2$-emitting area at the tip 
of the pillars.

 We note that the two ISM components traced by \brg  and \mh\, arise in adjacent but noticeably separate areas, and the local, apparent
line ratios are subject to significant projection effects. The projection effect is most severe in regions where the ionized gas envelops 
a molecular pillar viewed edge-on, such as in Area C.

\section{Photoionization Models} \label{S:cloudy_models}
\subsection{Motivations}
 Our main motivation to carry out numerical simulations is to explore a wide range of physical conditions in \HII regions and PDRs, and 
obtain emission line intensities as a function of physical conditions. The observed 
\brg and one single \mh\, emission provide very limited information, but with 
 numerical simulations, we are still able to learn something useful about the
physical conditions in 30 Dor.
Another motivation is to explore the issue of bright emission line contamination in the data. 
A series of \mh\, emission lines, hydrogen recombination lines, and He recombination lines are present in the 2$\micronm$ regime. We 
found evidence of additional emission line contamination (other than \brg) in the Ks continuum, affecting continuum 
subtraction in the \mh\, image (see \S~\ref{SSS:brightline}). 
We have empirically corrected for the contamination, and the estimated contamination level is in good agreement with similar type of 
observations in other \HII regions. Numerical simulations are helpful to investigate feasible origins of contamination and consequences.

\subsection{Model Parameters}\label{SS:parameters}
We generated models of simple \HII regions to study emission line intensities and physical conditions of molecular clouds in 30 Dor. 
First we used Starburst99 \citep{sb99} to generate ionizing continuum spectra of a massive coeval star cluster at 2 Myr age, because 
the age of R136 is $\lesssim$ 2 Myr \citep{dekoter98,massey98}.  The star cluster
is assumed to be massive enough to fully sample the initial mass function, which has exponents -1.3 and -2.3 between stellar mass boundaries 
of 0.1, 0.5, and 120 M$_\sun$. 
We employed the Geneva high mass-loss evolutionary tracks with 0.4 solar metallicity. The Geneva high mass-loss tracks are optimized for modeling atmospheres of high mass stars and are recommended by \citet{mm94}. We adopt Pauldrach/Hillier atmospheres and the LMC UV line library. The atmospheres include non-LTE and line-blanketing effects \citep{smith02} for O stars \citep{pauldrach01} and Wolf-Rayet stars \citep{hm98}.

Starburst99 output continuum spectra are fed into Cloudy 08.00\footnote{Calculations were performed with version 08.00 of Cloudy, last described by \citet{cloudy98}.} as the ionizing continuum of each simulated H II region. 
\citet{pellegrini10,pellegrini11} suggested that the inner 15 pc of 30 Dor lacks ionized gas and thus molecular clouds.  
 The ionization front near Area A identified in \citet{pellegrini10} 
is located at a projected distance of 30 pc, and materials behind the star cluster R136 are at a characteristic distance of 60 pc. 
We therefore set the inner radius of the simulated \HII regions 
at 30 pc and 60 pc with respect to the ionizing source\footnote{ Regions modeled at a projected distance $<$ 10 pc would require ionizing sources other than R136 \citep{pellegrini11}.}.
A wide density grid log(n$_{\rm H}$)=1 to 5 is incorporated in the calculations. 
We adopt the dust grain size distributions of the LMC  \citep{pellegrini11, weingartner01}, and Cloudy's default ISM abundances at 0.4 solar metallicity.
Each calculation stops when the cloud temperature drops to 30 K, well inside a PDR. 
Integrated line luminosities, \brg at 2.17$\micronm$, \HeI at 2.06$\micronm$, \mh\, 1--0 S(1) at 2.12$\micronm$, \mh\, 1--0 S(0) at 2.22$\micronm$, and \mh\, 2--1 S(1) at 2.25$\micronm$,
are calculated and recorded.

\begin{figure}
\begin{center}
\plotone{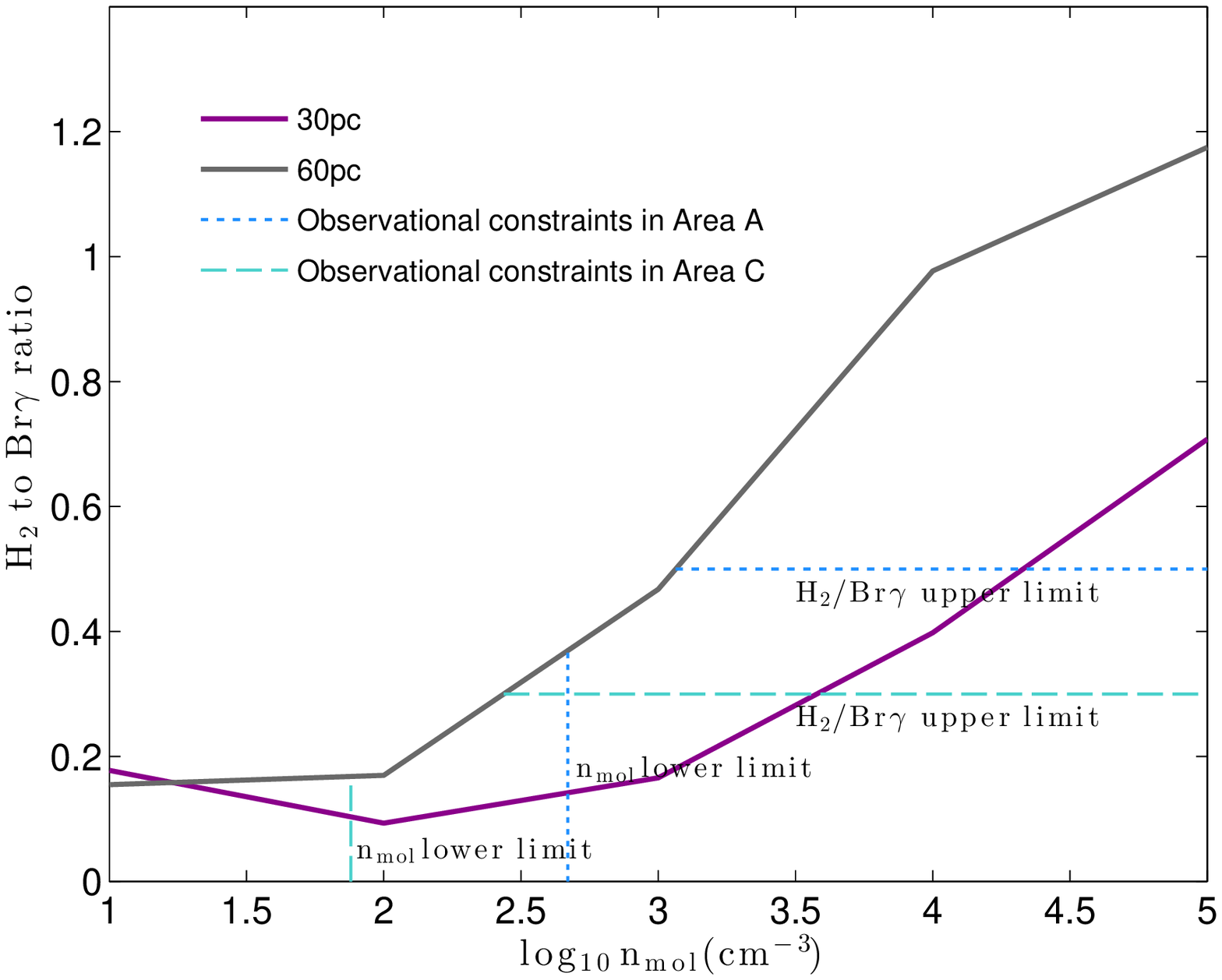}
\epsscale{1}
\caption{\mh\, 1--0 S(1) to \brg line ratios as a function of molecular gas density (n$\rm _{mol}$) and distance to the ionizing source.
 The purple line shows \mh\, to \brg line ratio computed at a distance of 30 pc, and the line ratio computed at 60 pc is shown in the gray line.
Observational constrains on \mh\, to \brg ratios and n$\rm _{mol}$  in Area A and C are marked by blue dotted and cyan dashed lines , respectively.
}
\label{fig:h2_to_brg_model}
\end{center}
\end{figure}

\section{Model Results and Discussion}\label{S:discussion}
With the model parameters indicated in \S~\ref{SS:parameters}, we first modeled \HII regions under perfect force balance
between thermal pressure of the cloud and incident radiation and wind pressure.
At the distance 30 pc and 60 pc away from the ionizing source, ionized gas density in the force-balanced \HII regions appears constant, however, 
in the PDR region the cloud pressure can become unrealistically high when the ionized gas density is high. 
This is because our simple models do not include turbulence and magnetic field pressure terms, which are 
important in supporting molecular clouds.The modeled regions 
thus form artificially high thermal pressure beyond the ionization front (IF) in order to meet the force balance criterium.
To resolve this problem, we turned to the constant density model. 
The \brg emission intensities calculated in both 
force balance and constant density models are consistent to within 1\%, therefore it is reassuring that the constant density models adequately 
represent the emission line spectra in the ionized gas, which is important in affecting the FUV spectra entered into PDR.
We generated tables of emission line intensities as a function of the density grid and interpolated the observed line ratios using such tables, 
to obtain plausible range of physical conditions in 30 Dor.

\subsection{Molecular Gas Density}\label{SS:h2_density}
We also modeled simple \HII regions with parameters described in \S~\ref{S:cloudy_models} and computed the \brg and \mh\, 1--0 S(1) emission intensities
as a function of gas density log$_{10}$\,n$_{\rm H}$ = 1 to 5 (Figure~\ref{fig:h2_to_brg_model}). 
The computed \mh\, to \brg line ratios at 30 pc and 60 pc are shown in purple and gray lines,
 and the dotted and dashed lines mark observational constraints of molecular gas densities and \mh\,/\brg ratios in Area A and C. 
 The lower limit of molecular gas densities is set by the ionized gas densities evaluated from the [SII] emission line doublet in \citet{pellegrini10} 
near the IFs at Area A and C. The ionized gas densities at those locations are found to be 10$^{2.7}$ cm$^{-3}$ and 10$^{1.9}$ cm$^{-3}$,
respectively. The upper limit of \mh\, to \brg line ratio in each area is set by our observations. 

 The \mh\, to \brg line ratio 
in most parts of Area A is $<$ 0.5, implying that the 
molecular gas density is $<$ 10$^{4}$ cm$^{-3}$ at 30 pc and $<$ 10$^{3}$ cm$^{-3}$ at 60 pc. 
The maximum \mh\, to \brg ratio in Area A is $\sim$0.5, and the corresponding molecular gas 
density would be $\sim$10$^{4.3}$ and 10$^{3.1}$ cm$^{-3}$ at a distance 30 pc and 60pc, respectively. 
The simple comparison of observed and modeled line ratios suggests that \mh\, in Area A is formed in relatively low density areas in
the PDR of 30 Dor, close to the surface
of the molecular cloud in conjunction to the IF, where ionized gas density is $\sim$ 10$^{2.7}$ cm$^{-3}$. 
This is consistent with our findings in \S\ref{S:results} that the observed \mh\, emission arises from the PDR near IFs. 

In the  Area C pillars, the maximum \mh\, to \brg line ratio measured is 0.3, where \mh\, emission intensity is also highest. 
 The projected distance of Area C is greater than 30 pc, we therefore refer to the 60 pc model. At the upper limit of 
\mh\, to \brg ratio of 0.3, 
the molecular gas density would be $\sim$10$^{2.4}$ cm$^{-3}$.
The Area C \mh\, emission is seen with a projected separation of 3 pc from its \brg envelope. 
With the ionized gas density 10$^{1.9}$ cm$^{-3}$, a depth of 3 pc would corresponds to A$_V$ $<$ 0.5.
 One expects molecular hydrogen to form inside a molecular cloud at depths A$_V \sim$ 0.13 \citep{vand88}, which further supports the notion that the observed \mh\, emission 
in Area C is formed inside the PDR but at a rather shallow depth. 
 This conclusion holds just as well for the other locations where we have identified \brg extending 2-3\,pc from an edge in the \mh\, emission. 
As noted in \S \ref{SS:h2_to_brg_ratios}, the apparent \mh/\brg line ratio determined on small scales is subject to projection effects because of this physical offset.

We can compare the observed \mh\, emission fluxes with PDR models such as \citet{sternberg89}(SD89) and \citet{black87}(BvD87), and constrain molecular gas densities in 
Area A, B, and C. The far UV radiation field $\chi$ in 30 Dor is $\sim$500$\chi_0$ (\citealt[][where $\chi_0$ \citep{draine78} relates to the \citet{habing68} field by a factor of 1.71]{pineda09,anderson14}), while SD89 and BvD87 predicted that the \mh\, 1--0 S(1) 
intensity is $\lesssim$2\% of the total \mh\, intensity at such radiation hardness. Adopting the average \mh\, emission intensities at Area A, B, and C as 2\% of 
the total \mh\, intensity, the corresponding molecular gas densities in those areas are $<$ 10$^{4}$ cm$^{-3}$, in coarse agreement with our Cloudy calculations.
To firmly constrain molecular cloud densities in the 30 Dor PDR, we will need multiple ro-vibrational transitions of \mh\, emission in followup studies.
With one \mh\, transition and \brg, 
nevertheless, our analysis suggests that molecular clouds associated with the observed \mh\, emission have densities $<$ 10$^4$ cm$^{-3}$.

\subsection{Fluorescence or Shock Excitation? Origin of the \mh\, 1--0 S(1) Line Emission}\label{SS:h2_origin}
NIR molecular hydrogen emission lines in HII regions can form either via (1) pure fluorescence excitation or (2) shock heating, and  
the best way to distinguish fluorescence from shock excitation is to analyze 
multiple transitions of ro-vibrational molecular hydrogen emission lines. 
Although such spectroscopic data do not yet exist, the \mh\, morphology, compared with CO morphology and \mh\,-to-\brg ratios, suggests no evidence of 
shock excitation.

As reported in \S~\ref{SS:morphology} and \S~\ref{SS:area_interest}, the \mh\, morphology generally correlates well with that of CO. \mh\, is likely to correlate poorly with CO emission in the case of shock excitation, which is often seen in star forming regions with active protostellar outflows. 
A good example is the Orion A giant molecular cloud \citep{davis09}. 

The \mh\,-to-\brg line ratio is another diagnostic to distinguish fluorescence from shock excitation. In shock-dominated regions, the line ratio is often found greater than unity \citep{puxley00,medling15}; while in massive star-forming regions, the \mh\,-to-\brg ratio is $<$ 0.6 \citep{joseph84,moorwood88,ra04,ra05,riffel10}. 
The reason behind the distinctive line ratios is fairly simple. Shocks cannot excite \brg emission, therefore in shock-dominated regions the \mh\,-to-\brg ratio will be high, regardless of viewing angle. 
On the other hand, UV radiation from massive stars excites \brg emission as well as \mh\,. \brg is often very bright, and when viewed face-on, one
naturally finds relatively low \mh\,-to-\brg ratios. 
In 30 Dor, the line ratio is no greater than 0.5 (see \S~\ref{SS:h2_to_brg_ratios}), which favors fluorescence as the dominant \mh\, excitation mechanism. 
We note that in regions viewed edge-on where \brg is seen spatially offset from \mh\,, such as the pillars in Area C, the \mh\,-to-\brg ratio will be very low. Nevertheless, the presence of 
both \brg and \mh\, excludes shock excitation, regardless of viewing angle.

\subsubsection{\mh\, and 8 $\micronm$ Emission Correlation}

The 8$\micronm$ emission adds additional information for diagnosing the origin of \mh\, emission in 30 Dor. 
Emission at 8$\micronm$ is largely dominated by Polycyclic Aromatic Hydrocarbon (PAH) emission, which 
is excited by far-UV radiation in PDRs. Shocks, on the other hand, will destroy the PAH molecules and suppress the emission \citep{flower03,micelotta10}. 
Therefore spatial and morphological correlations between the \mh\, and 8$\micronm$ emission provide hints 
 of the excitation mechanism of the observed \mh\, emission. Good morphological correlations between the two distributions
 would suggest that \mh\, emission is predominantly fluorescence excited, just like the 8$\micronm$ emission; poor morphological correlations would
otherwise imply that the FUV radiation is not a major excitation source of the observed \mh\, emission. 
The only excessive H2 emission related to shock activities is found near protostars at 4.5$\micronm$, e.g. \citet{cyganowski08} and \citet{lee13}.

\begin{figure}
\begin{center}
\plotone{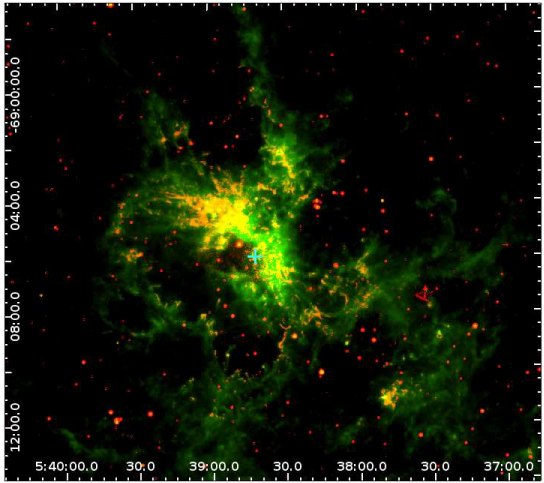}
\epsscale{1}
\caption[]{Spatial and morphological correlations between \mh\, and 8$\micronm$ emission in 30 Doradus.
\mh\, emission is shown in red, and the image is convolved to a 2.0$\arcsec$ resolution to match that in the 8$\micronm$ image.
The 8$\micronm$ emission is shown in green. R136 is marked by a cyan cross. 
}
\label{fig:h2_8um}
\end{center}
\end{figure}

\begin{figure}
\begin{center}
\plotone{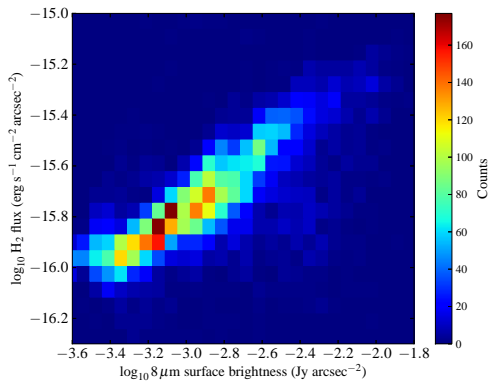}
\epsscale{1}
\caption[]{Surface brightness of \mh\, and 8$\micronm$ in Area A of 30 Doradus.
The color bar indicates the number of pixels in each surface brightness bin of \mh\, and 8$\micronm$ data.
The \mh\, and 8$\micronm$ intensities show an excellent correlation in this area, implying
that the \mh\, emission originated inside the PDRs. 
}
\label{fig:h2_8um_flux}
\end{center}
\end{figure}

We convolved the \mh\, image and Spitzer IRAC 8 $\micronm$ image to the same spatial resolution of 2.0$\arcsec$, and the superimposed 
image is shown in Figure~\ref{fig:h2_8um}. The \mh\, emission is displayed in red, while the 8$\micronm$ emission is in green. 
The overall morphologies of \mh\, and 8$\micronm$ emission correlate very well in the entire 30 Dor nebula, as well as in all three areas of interest A, B, and C. 
In Figure~\ref{fig:h2_8um_flux}, we plotted the surface brightness of \mh\, and 8$\micronm$ emission in Area A. The
\mh\, and 8$\micronm$ emission intensities also show a rather tight correlation.
The good morphological and intensity correlations between \mh\, and 8$\micronm$ emission strongly implies that the 
observed \mh\, emission is predominantly excited by the FUV radiation in the PDR. 

 In several places within the nebula,  especially Area C, \mh\, and 8 $\micronm$ emission highlight pillar structures.  These are enclosed by thicker layers of ionized gas traced by \brg (noted in \S\ref{S:results}), which in turn have sharp boundaries that presumably arise from the pressure of the x-ray emitting hot gas. An example is shown in a zoom-in figure in Figure~\ref{fig:3color-scale}. 
  In other locations, such as the finger in Area A, the 8 $\micronm$ emission shares an inner edge with the \brg emission, as though the grains responsible for this emission permeate the photoionized gas.  We intend to further pursue the physical interpretation of these correlations in a future paper.

\subsection{Bright Emission Line Contamination}
We use our models to further explore the bright emission line contamination issue in our \mh\, image. 
\brg at 2.17$\micronm$ and \mh\, 1--0 S(1) at 2.12$\micronm$ are present in the modeled continuum spectra and
are very close to the center of the Ks band filter, as expected. 
 In addition,  there are  \HeI 2$^{~1}$P--2$^{~1}$S emission line at 2.06$\micronm$, \mh\, 1--0 S(2) at 2.03$\micronm$, 
 \mh\, 1--0 S(0) at 2.22$\micronm$, and \mh\, 2--1 S(1) at 2.25$\micronm$. Table~\ref{tab:modeled_lines} summarizes the 
emission lines in the Ks band.

The  \mh\, 1--0 S(0), \mh\, 1--0 S(2), and \mh\, 2--1 S(1) lines are negligible, for they contribute $<$ 10\% of the \brg flux in the Ks band. 
The \HeI line is the most plausible source of continuum contamination in our data.
 The calculated line ratios \HeI to \brg at densities $<$ 10$^4$ cm$^{-3}$ are 0.6 to 0.7,
which suggests that the \HeI line will contribute 60\% to 70\% of the \brg intensity to
 the continuum level, in agreement with the empirically evaluated 61\% level in \S~\ref{SSS:brightline} and the 70\% level in M16 (L00).  
 Since \brg emission is 7\% of the Ks continuum, the \HeI contamination in the filter will be up to 4\%,
 assuming the distribution is the same.

 \begin{deluxetable}{lc}
\tablewidth{0pt}
\tablecaption{Emission lines within the Ks bandpass}
\tablehead{
\colhead{Line} &
\colhead{Wavelength} 
}
\startdata
\brg & 2.17$\micronm$ \\  
\HeI 2$^{~1}$P--2$^{~1}$S & 2.06$\micronm$\\
\mh\, 1--0 S(2) & 2.03$\micronm$ \\
\mh\, 1--0 S(1) & 2.12$\micronm$ \\ 
\mh\, 1--0 S(0) & 2.22$\micronm$ \\
\mh\, 2--1 S(1) & 2.25$\micronm$
\enddata
\label{tab:modeled_lines}
\end{deluxetable}

\section{Summary}\label{S:summary}
We present the first and fully calibrated \mh\, 1--0 S(1) emission image of the entire 30 Doradus nebula,
as well as a \brg image. In the data reduction process, we confirmed \brg and \HeI emission line contamination 
in the Ks continuum via empirical analysis and Cloudy simulations. The error in contamination-corrected \mh\, images is estimated to be $\sim$ 4\%. 

The overall morphology of \mh\, correlates well with \brg and CO emission, implying that 
the observed \mh\, originates from the PDRs in 30 Doradus. 
The brightest H$_2$-emitting areas (Area A and Area B) are PDRs viewed face-on located behind the ionizing source R136, and 
the warm molecular clouds traced by \mh\, appear to be clumpy.
Those regions also trace the CO morphology well, further indicating that these PDRs 
are face-on, with layers of ionized gas, warm molecular gas, and cold molecular clouds. 
 Discontinuity of \mh\, and \brg morphology is found at the outer shells and pillar features (such as Area C) in 30 Dor , 
where the \mh\, pillars are encompassed by \brg envelops of sharp boundaries.
This suggests that 
we are viewing the the shells and pillars of the \HII region edge-on.
The mean projected separation between the \brg envelope and \mh\, clumps is 3 pc (A$_V$ $<$ 0.7), indicating that \mh\, emission is formed 
in the PDRs close to the surface of the molecular clouds.
The density of \mh-emitting gas is inferred from the observed \mh-to-\brg line ratios, the Cloudy model results, as well as theoretical predictions of 
\mh\, emission in PDRs. The molecular gas 
density is estimated to be $<$ 10$^4$ cm$^{-3}$.
Low \mh-to-\brg line ratios ($<$ 0.5), as well as good morphological correlations between the \mh\, and 8$\micronm$ emission,
implying again that the observed \mh\, emission is excited by FUV radiation.

While it requires multiple transitions of ro-vibrational \mh\, lines to constrain the excitation mechanisms and physical parameters of the ISM, such data do not yet exist. Our imaging observations suggest that the observed \mh\, emission likely arises from a lower density layer of the PDR 
near IFs in 30 Doradus. We found no sign of shock-excited \mh\, emission, and all indications were consistent with fluorescent excitation.

\acknowledgements
We thank the anonymous referee for detailed and constructive comments. We also thank all CTIO 4-meter Blanco Telescope staff for their assistance during the observing runs. 
SCCY acknowledges help on data reduction from Ron Probst, Mark Dickinson, and Robert Swaters. SCCY's research is supported by a NAOJ-Subaru Telescope Fellowship and a University of Toronto Fellowship. CDM's research is supported by an NSERC Discovery grant.  We are also pleased to thank Lee Armus, Tim Heckman, Norman Murray, and Peter Martin for useful discussions.

\end{document}